\documentclass[a4paper,11pt]{article}
\usepackage{pos}
\usepackage{cleveref}
\usepackage{graphicx}
\usepackage{subfig}
\usepackage{booktabs}
\usepackage{makecell}

\title{Two-loop amplitude for $t\bar{t}W$ production at hadron colliders in the leading colour approximation}
\ShortTitle{Two-loop amplitude for $t\bar{t}W$ production}

\author*[a]{Mattia Pozzoli}

\affiliation[a]{Dipartimento di Fisica e Astronomia, Universit\`{a}	 di Bologna, \\
INFN, Sezione di Bologna, \\
via Irnerio 46, I-40126 Bologna, Italy}

\emailAdd{mattia.pozzoli@unibo.it}

\abstract{In this contribution I present the first exact calculation of the leading-colour two-loop QCD amplitude for the associated production of a top-anti-top pair and a W boson. I discuss strategies to address the complexity of the computation, which involves complicated analytic structures, such as nested square roots, elliptic functions, and expressions with a high degree of algebraic complexity. The final result is expressed in terms of a set of special functions, which are evaluated using the method of differential equations, and rational coefficients, evaluated via finite field techniques.}

\FullConference{Loops and Legs in Quantum Field Theory (LL2026)\\
12-17, April, 2026\\
Bayreuth, Germany\\}

\begin{document}
\maketitle

\section{Introduction}
We are interested in the associated production of a top-antitop pair with a $W$ boson ($t\bar{t}W$) for various reasons. Firstly, this process is relevant to searches for physics beyond the Standard Model~\cite{Buckley:2015lku,Dror:2015nkp,BessidskaiaBylund:2016jvp}. Secondly, it constitutes a significant background for the production of a top-antitop pair with a Higgs boson ($t\bar{t}H$) and for four-top production ($t\bar{t}t\bar{t}$).

From the experimental point of view, the most precise measurements for the $t\bar{t}W$ total cross section are reported in~\cite{CMS:2022tkv,ATLAS:2024moy}, and some differential measurements are also available~\cite{ATLAS:2024moy}. Crucially, these measurements are in tension with the theoretical predictions, calling for increased precision on the theory side. In this context, results for the NNLO QCD predictions have been obtained in~\cite{Buonocore:2023ljm}, approximating the two-loop amplitude by combining the soft-$W$ approximation~\cite{Catani:2022mfv,Barnreuther:2013qvf} and the procedure of massification~\cite{Penin:2005eh,Mitov:2006xs,Becher:2007cu}. In the bulk of the phase space, the approximation is expected to correctly capture the exact amplitude within the corresponding systematic uncertainties. Thus, it is adequate for the computation of the total cross section, where the impact of the double virtual has been estimated to be around 6-7 \%. However, the computation of the exact two-loop amplitude for this process remains necessary both to validate the approximation and, potentially, to compute differential distributions.

To this end, the one-loop amplitude to higher orders in the dimensional regulator $\varepsilon$ has been computed in~\cite{Becchetti:2025osw}. In these proceedings, I present the numerical evaluation of the two-loop amplitude in the leading colour approximation. The presence of seven kinematic variables and internal massive propagators makes the computation particularly challenging. On the one hand, massive virtual propagators lead in this case to the appearance of transcendental functions beyond the polylogarithmic case, associated with elliptic curves. On the other hand, the presence of many variables in the problem inevitably yields algebraically complicated expressions. In this computation, these two sources of complexity are intertwined. As was already observed in the computation of the two-loop Feynman integrals relevant for the leading-colour part of the amplitude~\cite{Becchetti:2025qlu}, gigantic algebraic expressions appear in the differential equations (DEs)~\cite{Barucchi:1973zm,Kotikov:1990kg,Kotikov:1991hm,Gehrmann:1999as,Bern:1993kr,Henn:2013pwa} satisfied by the master integrals (MIs).

\section{Two-loop form factors}
We assign the momenta of the particles as follows:
\begin{equation}
    \bar{u}(p_1) + d(p_2) + \bar{t}(p_3) + t(p_4) + W^+ (p_5) \to 0,
    \label{eq:ScatteringProcess}
\end{equation}
where the external particles are on-shell
\begin{equation}
    \label{eq:on_shell}
    p_1^2 = p_2^2 = 0,\quad p_3^2 = p_4^2 = m_t^2, \quad p_5^2 = m_W^2.
\end{equation}
and we impose momentum conservation as
\begin{equation}
    \label{eq:momentum_conservation}
    p_1+p_2+p_3+p_4+p_5=0.
\end{equation}
We describe the kinematics of the process by seven Lorentz invariants which we choose as five Mandelstam variables $s_{ij} = (p_i +p_j)^2$ and the masses of the top quark and of the $W$ boson
\begin{equation}
    \vec{x} := 
    \left\{s_{13} , s_{34} , s_{24} , s_{25} , s_{15}, m_t^2 , m_W^2 \right\}.
\end{equation}
In order to regulate the divergences of the Feynman integrals we use dimensional regularisation, working in $D=4-2 \varepsilon$ space-time dimensions.

The amplitude admits an expansion in the number of colours $N_c=3$ and of light-quark flavours $N_f=5$
\begin{equation}
\mathcal{A}^{(2)} = N_c^2 \mathcal{A}^{(2,N_c^2)}+N_c N_f \mathcal{A}^{(2,N_c N_f)}+N_f^2 \mathcal{A}^{(2,N_f^2)} + \mathcal{O}(N_c),
\label{eq:leading_colour}
\end{equation}
where each term is individually gauge-invariant. We work in the generalised leading colour approximation, retaining only the terms proportional to $N_c^2, N_c N_f$ and $N_f^2$. This reduces the number of Feynman diagrams from 722 to 210, all of which are planar and involve at most two massive propagators.

Following~\cite{Peraro:2019cjj,Peraro:2020sfm}, we work in the 't Hooft-Veltman (tHV) scheme~\cite{tHooft:1972tcz} to further decompose the leading colour amplitude into the same 24 independent tensor structures $T_i$ of~\cite{Becchetti:2025osw} as
\begin{equation}
\mathcal{A} = \sum_{i=1}^{24} F^{(i)} T_i,
\label{eq:ff_decomposition}
\end{equation}
where the form factors $F^{(i)}$ are functions of the kinematics. These are expressed as linear combinations of 8959 Feynman integrals with rational coefficients.

\section{Calculation of the Feynman integrals}
\begin{figure}[t]
\begin{center}
\subfloat[Family $A$: 17 MIs]{\includegraphics[scale=0.25]{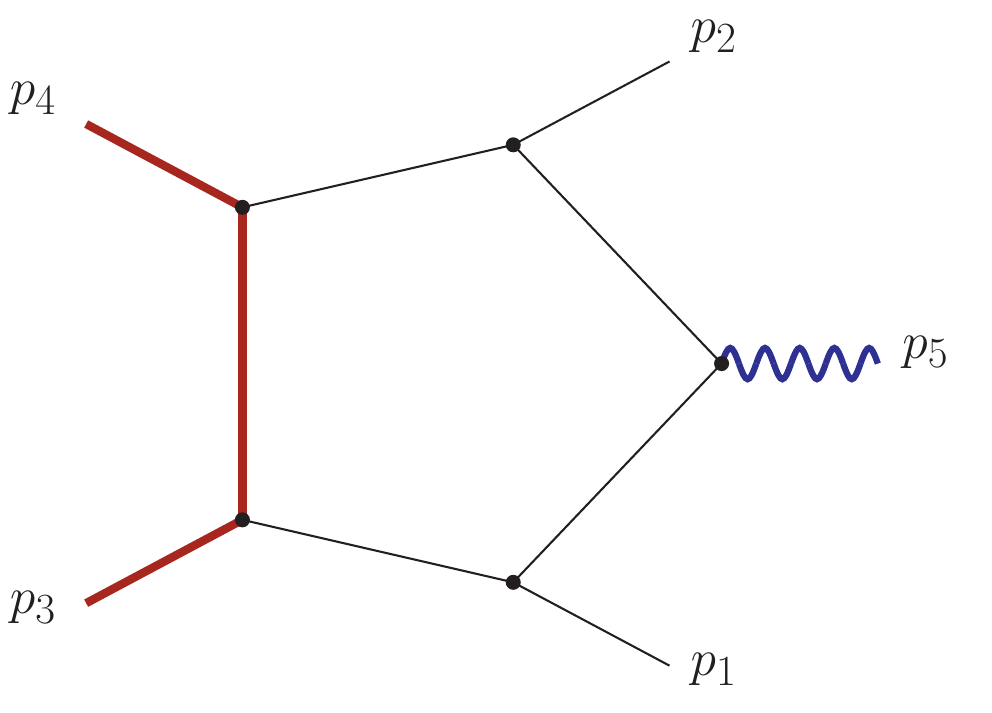}\label{fig:famA}}\quad
\subfloat[Family $F_1$: 141 MIs]{\includegraphics[scale=0.25]{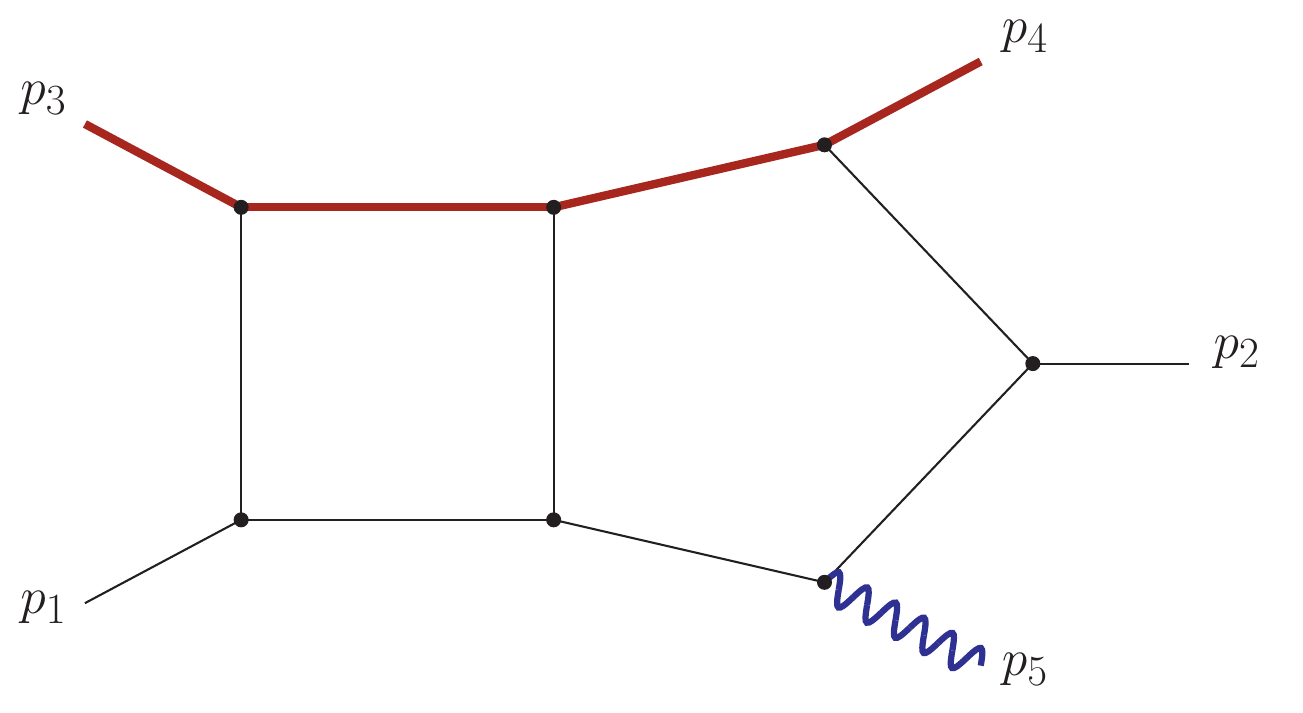}\label{fig:famF1}}\\
\subfloat[Family $F_2$: 122 MIs]{\includegraphics[scale=0.25]{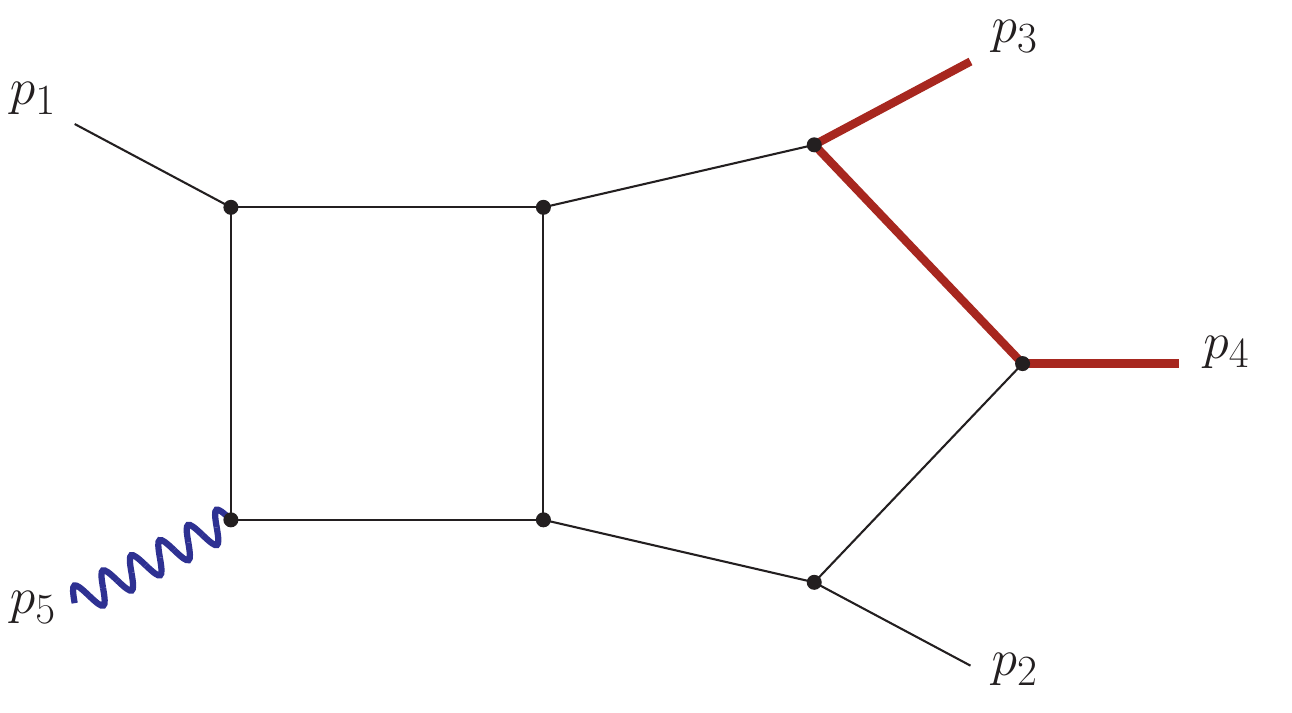}\label{fig:famF2}}\quad
\subfloat[Family $F_3$: 131 MIs]{\includegraphics[scale=0.25]{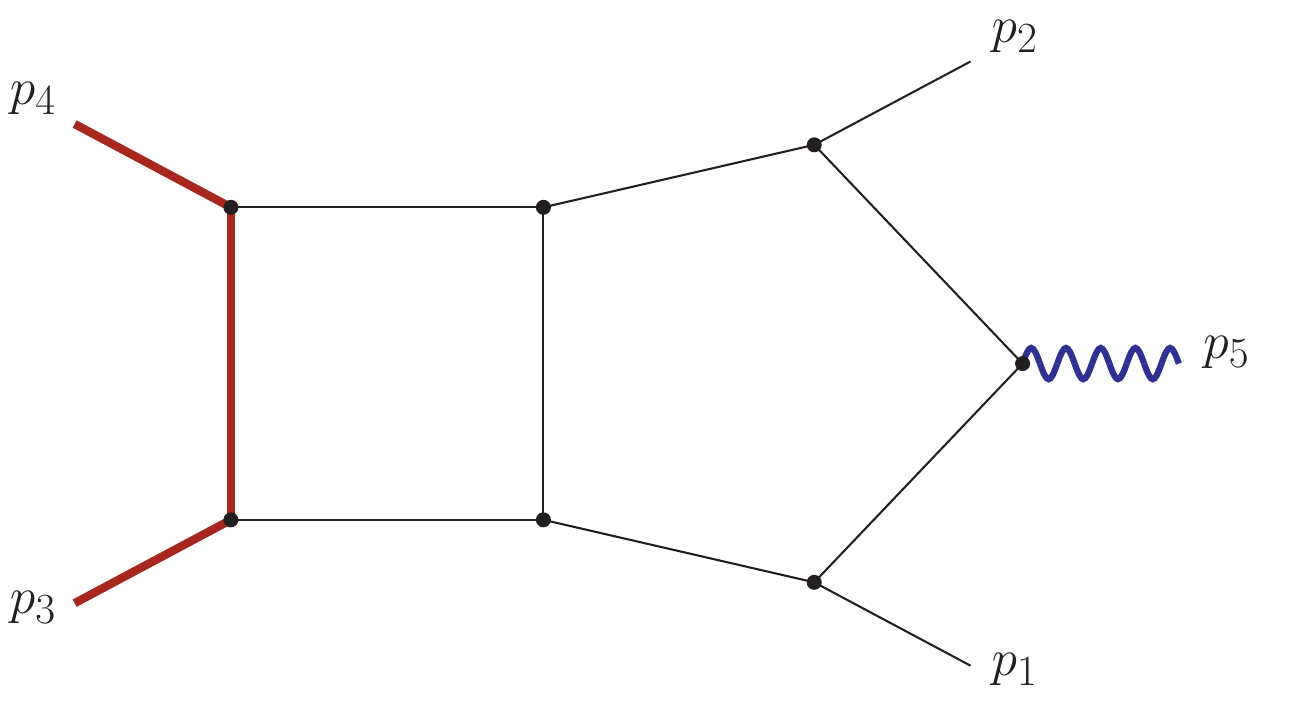}\label{fig:famF3}}
\caption{One-loop (\cref{fig:famA}) and two-loop (\cref{fig:famF1,fig:famF2,fig:famF3}) integral families contributing to the leading-colour amplitude. Thin black lines denote massless particles, thick red lines indicate the top quarks and the curly blue line is the $W$ boson.}
\label{fig:allFams}
\end{center}
\end{figure}
We group Feynman integrals into integral families, defined by their set of denominators, such that any integral is defined by specifying a set of exponents as
\begin{equation}
G_{a_1,\dots, a_s} = \int \left( \prod_{l=1}^L \frac{\mathrm{d}^{D}k_l e^{\varepsilon \gamma_E}}{\mathrm{i} \pi^{\frac{D}{2}}} \right) \frac{1}{\prod_{i=1}^s D_i^{a_i}}, \quad (a_1, \dots, a_s) \in \mathbb{Z}^s,
\label{eq:integral_family}
\end{equation}
where $L$ is the number of loops and $s=5$ for one-loop integrals and $s=11$ for two-loop families. Up to permutations, the Feynman integrals belong to the three irreducible two-loop families that were studied in~\cite{Becchetti:2025qlu}, whose top-sector graphs are shown in \cref{fig:famF1,fig:famF2,fig:famF3}. Additionally, some integrals are products of two lower loop integrals, belonging to the family in \cref{fig:famA} that was computed in~\cite{Becchetti:2025osw}.

It is known that the elements of an integral family satisfy linear relations, such as integration-by-parts identities (IBPs)~\cite{Tkachov:1981wb,Chetyrkin:1981qh,Laporta:2001dd}, which allow us to write the Feynman integrals as linear combinations of a finite basis $\vec{I}$ of master integrals (MIs). We use \textsc{NeatIBP}~\cite{Wu:2023upw} to generate the IBPs and \textsc{FiniteFlow}~\cite{Peraro:2019svx} to solve them, reducing the amplitude to 330 MIs.

\subsection{Differential equations for the master integrals}
From the IBPs, we derive a system of linear differential equations (DEs)~\cite{Barucchi:1973zm,Kotikov:1990kg,Kotikov:1991hm,Gehrmann:1999as,Bern:1993kr,Henn:2013pwa} for the MIs as
\begin{equation}
\forall \ \xi \in \vec{x} : \partial_{\xi} \vec{I} (\vec{x};\varepsilon)=B_{\xi} (\vec{x};\varepsilon)\cdot  \vec{I}(\vec{x};\varepsilon),
\label{eq:DEs}
\end{equation}
where the matrices $B_{\xi}$ generally involve complicated rational functions that mix the dependence on $\vec{x}$ and $\varepsilon$. This raises the question of which basis of MIs to pick, as a good choice can greatly simplify the DEs. Ideally, we would like to obtain DEs in the so-called canonical form~\cite{Henn:2013pwa}
\begin{equation}
\mathrm{d} \vec{I}(\vec{x}; \varepsilon) = \varepsilon \ \mathrm{d} \tilde{A} (\vec{x}) \cdot \vec{I} (\vec{x};\varepsilon),
\label{eq:canonical_basis}
\end{equation}
characterised by the factorisation of the dependence on $\varepsilon$ and by the fact that the connection matrix $\tilde{A}$ involves differential forms with locally at most simple poles. In the best understood case, when the integrals integrate to polylogarithms, the connection matrix is expressed in terms of logarithmic differential forms as
\begin{equation}
\mathrm{d} \tilde{A}(\vec{x}) = \sum_i a^{(i)} \ \mathrm{d} \log W_i (\vec{x}),
\label{eq:dlog_connection}
\end{equation}
where the $a^{(i)}$ are matrices of rational numbers and the algebraic functions of the kinematics $W_i (\vec{x})$ are called letters.

In the polylogarithmic case of \cref{eq:dlog_connection}, we can construct MIs satisfying canonical DEs using the method of integrand analysis~\cite{Arkani-Hamed:2010pyv,Henn:2020lye}, based on the conjecture that integrands with at most simple poles and a constant leading singularity lead to canonical Feynman integrals. Practically, we look for integrals that in some parametrisation look like
\begin{equation}
I \propto R(\vec{x}) \int \mathrm{d}\log(f_1(\vec{z})) \wedge \mathrm{d}\log(f_2(\vec{z})) \wedge \dots \wedge \mathrm{d}\log(f_n(\vec{z})),
\label{eq:integrand_analysis}
\end{equation}
where the $f_i(\vec{z})$ are algebraic functions of the integration variables $\vec{z}$ and $R(\vec{x})$ is the leading singularity. The conjecture then tells us that $\frac{1}{R(\vec{x})} I$ should satisfy canonical DEs. Most of the integrals appearing in the families in \cref{fig:famF1,fig:famF2,fig:famF3} fall under this case, and we construct a basis of canonical MIs for them.

\begin{figure}[t]
\begin{center}
\subfloat[]{\includegraphics[scale=0.25]{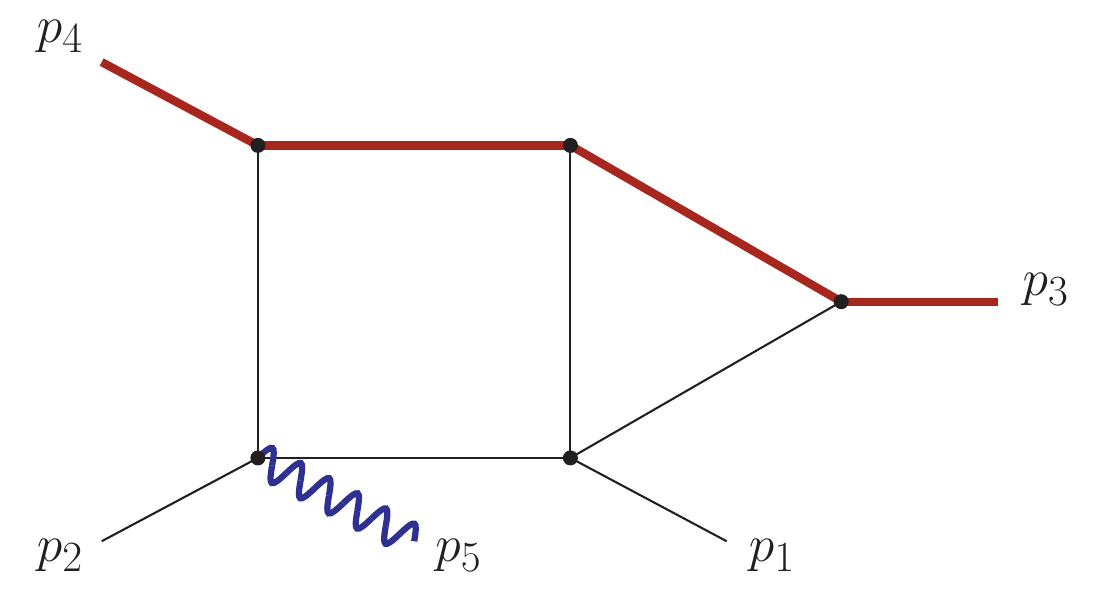}\label{fig:ttj_elliptic}}\quad
\subfloat[]{\includegraphics[scale=0.25]{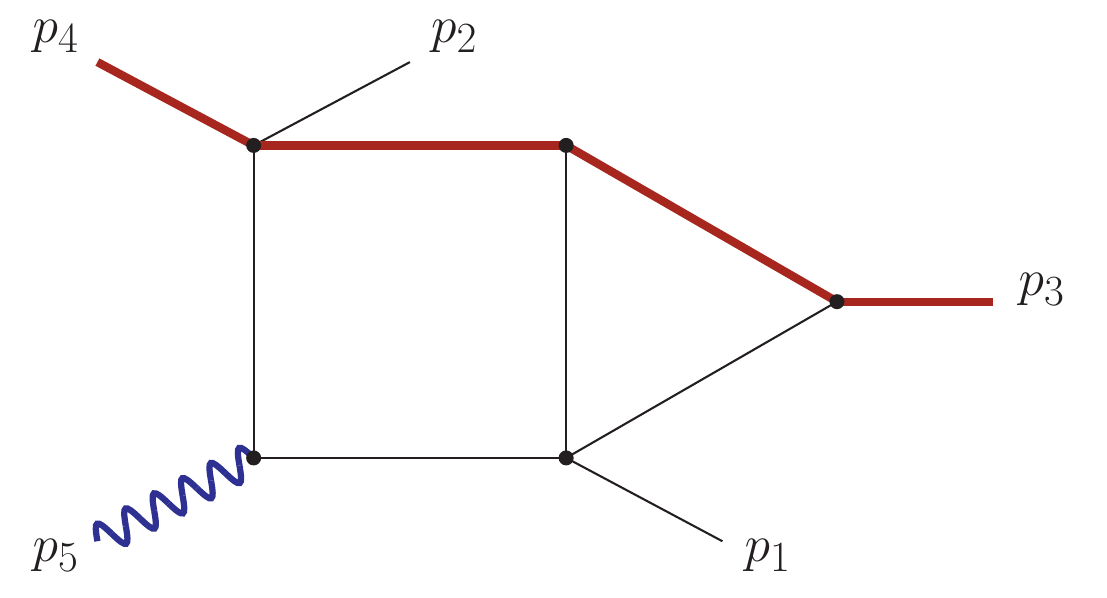}\label{fig:elliptic_simple}}\\
\subfloat[]{\includegraphics[scale=0.25]{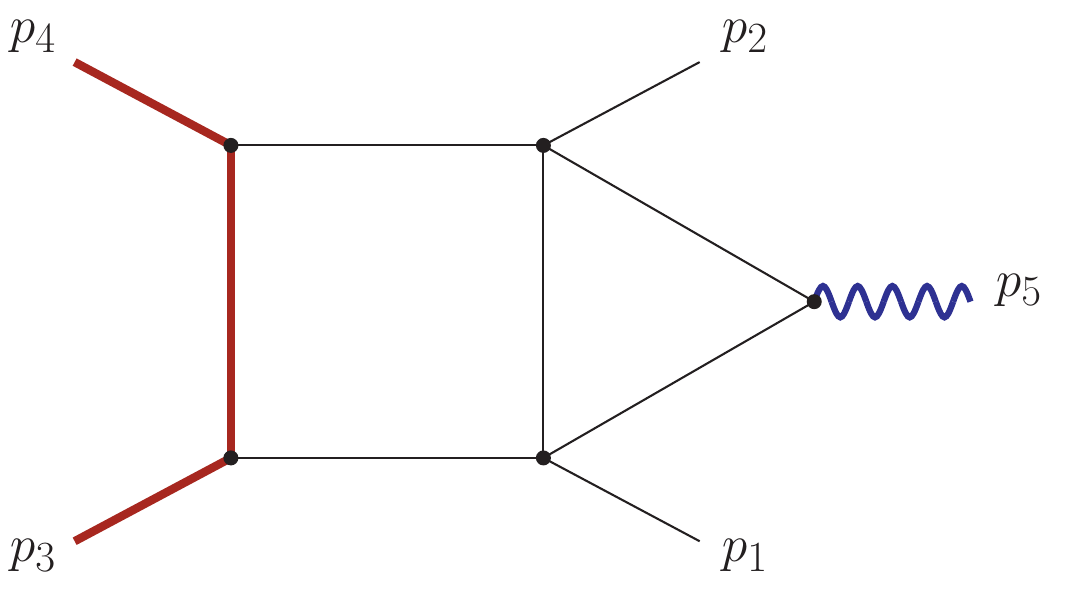}\label{fig:elliptic_monster}}\quad
\subfloat[]{\includegraphics[scale=0.25]{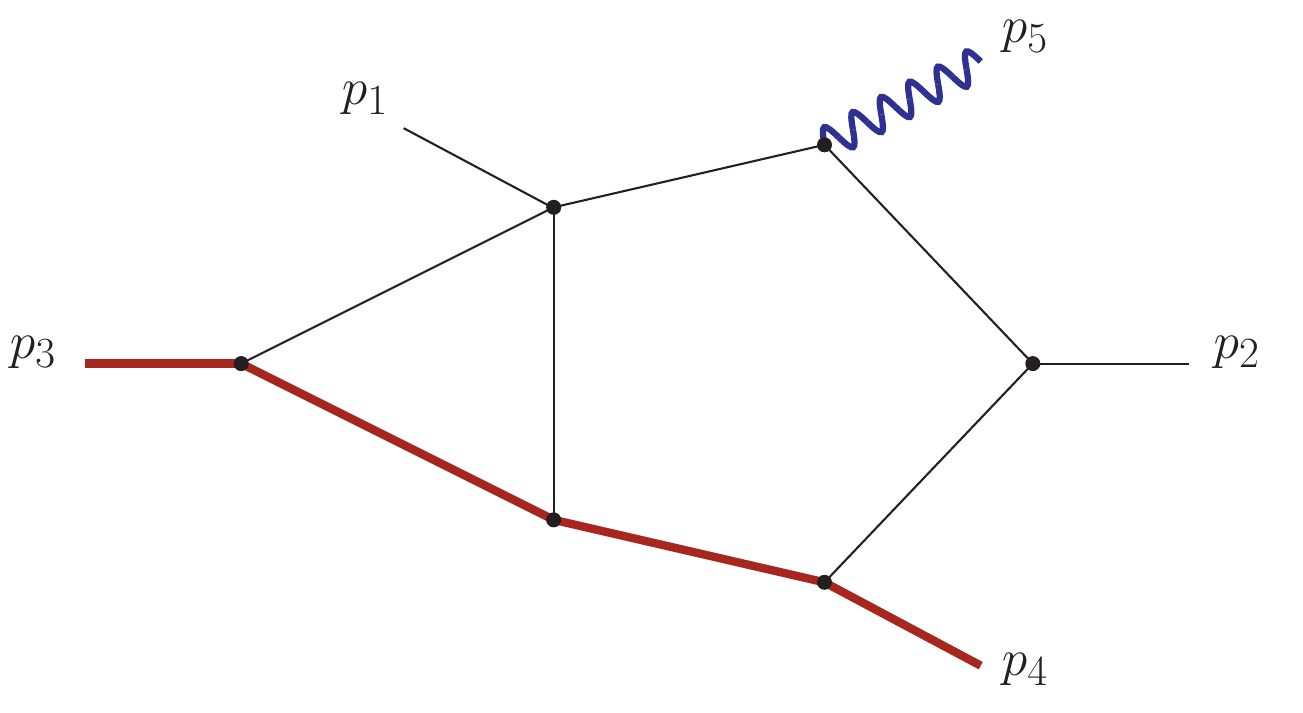}\label{fig:nested_root}}
\caption{Graphs of the sectors associated with the elliptic curves (\cref{fig:ttj_elliptic,fig:elliptic_simple,fig:elliptic_monster}) and with the nested square root (\cref{fig:nested_root}).}
\end{center}
\end{figure}

However, \cref{eq:integrand_analysis} could reveal also differential forms more complicated than the logarithmic ones, associated for example with an elliptic curve
\begin{equation}
\frac{\mathrm{d}z}{\sqrt{\mathcal{P}_4(z)}}, \quad \mathrm{with} \quad \mathcal{P}_4(z) = (z-a_1) (z-a_2) (z-a_3) (z-a_4),
\label{eq:elliptic_curve}
\end{equation}
where the roots $a_1, \dots, a_4$ of the polynomial $\mathcal{P}_4(z)$ are non-degenerate. Strategies to construct canonical integrals also in these cases are an active field of study in the community. Nevertheless, for processes involving five external legs the first canonical DEs have been obtained only recently for $t \bar{t}j$ production~\cite{Becchetti:2025oyb}. In the calculation of the two-loop MIs, we find three sectors associated with an elliptic curve, and an additional source of analytic complexity is the presence of a nested square root in the sector in \cref{fig:nested_root}. 

The two sectors in \cref{fig:ttj_elliptic,fig:elliptic_simple} have a four-point kinematics and are comparable with integrals already known in the literature. The sector in \cref{fig:elliptic_monster}, on the other hand, has a five-point kinematics, yielding a leap in the algebraic complexity of the elliptic curve compared to the other sectors. For instance, the discriminant of the curve involves a degree 14 irreducible polynomial, which appears both in the denominators of the differential equation and in the Landau discriminant, as we checked using \textsc{SOFIA}~\cite{Caron-Huot:2024brh,Correia:2025wtb}.

Because of this algebraic complexity, constructing a canonical basis for this sector would be challenging even with the known methods, and we thus refrain from doing so. Instead, we follow the strategy of~\cite{Badger:2024fgb} to choose MIs that satisfy DEs of the form
\begin{equation}
\begin{split}
\mathrm{d} \vec{I}(\vec{x};\varepsilon) &= \mathrm{d} A (\vec{x};\varepsilon) \cdot \vec{I} (\vec{x};\varepsilon),\\
\mathrm{d} A(\vec{x};\varepsilon) &= \sum_{k=0}^2 \varepsilon^k \big[ \sum_{\alpha} c_{k\alpha} \mathrm{d} \log(W_{\alpha}(\vec{x}))+\sum_{\beta} d_{k\beta} \omega_{\beta}(\vec{x}) \big],
\end{split}
\label{eq:des_structure}
\end{equation}
where the non-logarithmic one-forms $\omega_\beta$ take the general form
\begin{equation}
\omega_\beta (\vec{x}) = \sum_{i=1}^7 c_i (\vec{x}) \ \mathrm{d} x_i,
\label{eq:omegabetas}
\end{equation}
and we chose them to be linearly independent and such that the polynomial degrees are minimised. The entries of \cref{eq:des_structure} involving only polylogarithmic integrals are $\varepsilon$-factorised, and contain only logarithmic one-forms.

\section{Evaluation of the amplitude}
In order to compute the amplitude, we do not need the full dependence of the integrals on $\varepsilon$, but only the first terms of their Laurent expansion around $\varepsilon=0$
\begin{equation}
I_j(\vec{x}; \varepsilon) =  \sum_{w=0}^{w_\mathrm{max}}  \varepsilon^w \ I_j^{(w)}(\vec{x}).
\label{eq:Laurent_expansion}
\end{equation}
For the two-loop amplitude, we only need to compute the master integral coefficients $I_j^{(w)}$ up to order $w_\mathrm{max}=4$. The master integral coefficients $I_j^{(w)}$ are not all independent, and this redundancy leads to complications in the representation of the amplitude and to a slower numerical evaluation.

In the polylogarithmic case, there are well-known techniques~\cite{Gehrmann:2018yef,Chicherin:2020oor,Chicherin:2021dyp} that allow us to determine a basis of algebraically independent special functions $\{f_k^{(w)} \} \subset \{ I_j^{(w)} \}$, such that any master integral coefficient is expressed as a graded polynomial in the special functions and some transcendental constants, conjectured to be only the zeta-values $\zeta_n$~\cite{Abreu:2023rco}. In order to extend the techniques to our case, we follow the approach of~\cite{Badger:2024dxo}. By choosing finite elliptic integrals (i.e.~$I_j^{(w)} = 0$ for $w \leq 3$ for all integrals in the elliptic sectors), one can show that the master integral coefficients up to order three in the $\varepsilon$-expansion satisfy canonical DEs with a connection matrix of the form of \cref{eq:dlog_connection}, and we can thus apply the techniques that are valid in the polylogarithmic case. The presence of elliptic integrals then manifests itself at order $w=4$ in the form of a few additional special functions that we label as $f_k^{(4,*)}$.

\begin{table}[t]
\begin{center}
\begin{tabular}{ccccccc}
\toprule
 & $f_k^{(1)}$ & $f_k^{(2)}$ & $f_k^{(3)}$ & $f_k^{(4)}$ & $f_k^{(4,*)}$ & Total\\
\midrule
\text{Form factors} & 7 & 12 & 63 & 212 & 29 & 323\\
\text{Finite remainder} & 7 & 12 & 63 & 187 & 29 & 298\\
\bottomrule
\end{tabular}
\caption{Number of special functions at each order appearing in the form factors and in the finite remainder.}\label{tab:special}
\end{center}
\end{table}

Expressing the amplitude in terms of special functions allows us to subtract the poles analytically, because the special functions stemming from the polylogarithmic MIs are algebraically independent by construction and the functions $f_k^{(4,*)}$ only appear in the finite part. In \cref{tab:special} we report the number of special functions appearing in the amplitude. One can see that the total number of special functions is lower than the number of MIs, and that some of the functions drop out from the finite remainder, i.e.~the renormalised and infrared subtracted amplitude.

\begin{table}[t]
\begin{center}
\begin{tabular}{cccccc}
\toprule
 & Square & Rational & Algebraic & All & Non-logarithmic\\
  & roots & letters & letters & letters & one-forms\\
\midrule
MIs & 13 & 65 & 61 & 126 & 220\\
Special functions & 13 & 52 & 48 & 100 & 138\\
\bottomrule
\end{tabular}
\caption{Analytic structures appearing in the differential equations for the MIs and for the special functions.}\label{tab:analytic_structures}
\end{center}
\end{table}

Like the MIs, the special functions satisfy linear differential equations, which however only depend on the kinematics and not on $\varepsilon$. Additionally, these DEs are sparser, and involve fewer analytic structures than the DEs satisfied by the MIs, as shown in \cref{tab:analytic_structures}. This is due to the fact that these structures would only appear at higher orders in the $\varepsilon$-expansion. For these reasons, evaluating the special functions is less time-consuming than evaluating the MIs. We solve the DEs using generalised series expansions~\cite{Pozzorini:2005ff,Moriello:2019yhu}, as implemented in the DE solver of \textsc{AMFlow}~\cite{Liu:2022chg}. Due to the algebraic complexity of the expressions involved, the evaluation is a bottleneck and it takes order of one hour per phase-space point.

As for the rational coefficients, their analytic reconstruction remains challenging due to their algebraic complexity. We thus evaluate them at each phase space point following the approach described in~\cite{Peraro:2019okx}: we reconstruct their exact value at rationalised phase space points from numerical evaluations over finite fields~\cite{vonManteuffel:2014ixa,Peraro:2016wsq} using \textsc{FiniteFlow}~\cite{Peraro:2019svx}. The number of needed primes, and hence the evaluation time, depends both on the polynomial degree of the coefficients and on the rationalisation precision of the invariants at the phase space point. In this context, the representation in terms of special functions helps improving the efficiency, since it reduces the degree of the coefficients. For instance, for the points we evaluated with our setup, we needed a maximum of 400 primes. Evaluating the rational coefficients of the $\varepsilon$-expanded MIs would have instead required 608 primes. As an additional optimisation, we use numerical samples over finite fields to determine linear relations between the rational coefficients~\cite{Badger:2021nhg}. This allows us to express the amplitude in terms of a linearly independent set of coefficients, which we choose to be the ones that require fewer primes for the reconstruction. In general, the evaluation of all coefficients at a single phase space point requires $\mathcal{O}$(10 minutes), and it is therefore not a bottleneck compared to the evaluation of the special functions.

\section{Conclusion}
I presented the numerical evaluation of the two-loop amplitude for $pp \to t\bar{t}W$ in the generalised leading colour limit, addressing the complications due to the presence of elliptic curves and nested square roots, and the considerable algebraic complexity of the expressions involved. The amplitude is expressed in terms of a possibly over-complete basis of special functions with rational coefficients. We evaluate the former through generalised series expansions, and we reconstruct the latter from evaluations over finite fields. This allows us to evaluate the finite remainder at a phase space point in the order of one hour. We employed our framework to produce an interpolation grid for the finite remainder, which we used to obtain results for the NNLO QCD inclusive cross section~\cite{Becchetti:2026awn}.

\acknowledgments

I would like to thank my collaborators M. Becchetti, D. Canko, X. Chen, V. Chestnov, M. Delto, S. Ditsch, M. Grazzini, S. Kallweit, T. Peraro, C. Savoini, L. Tancredi and S. Zoia for all the great work that lead to the results I presented in these proceedings. This work was supported by the European Research Council (ERC) under the European Union's Horizon Europe research and innovation program grant agreement 101040760, \textit{High-precision multi-leg Higgs and top physics with finite fields} (ERC Starting Grant \emph{FFHiggsTop}).

\bibliographystyle{JHEP}
\bibliography{biblio}

\end{document}